\documentclass[12pt]{article}

\usepackage{newtxtext,newtxmath}

\usepackage{graphicx}

\usepackage[letterpaper,margin=1in]{geometry}

\renewenvironment{abstract}
	{\quotation}
	{\endquotation}

\date{}

\makeatletter
\renewcommand{\fnum@figure}{\textbf{Figure \thefigure}}
\renewcommand{\fnum@table}{\textbf{Table \thetable}}
\makeatother

\usepackage{scicite}

\usepackage{url}

\def\scititle{
Ionospheric Electron Heat Flow Modulates Planetary Ambipolar Electric Fields
}
\title{\bfseries \boldmath \scititle}

\author{
	Liangliang Yuan$^{1\ast}$\and
	Shuanggen Jin$^{1,2\ast}$\and
	\small$^{1}$Shanghai Astronomical Observatory, Chinese Academy of Sciences, Shanghai \& 200030, China.\and
	\small$^{2}$School of Surveying and Land Information Engineering, Henan Polytechnic University, Jiaozuo \& 454003, China.\and
	\small$^\ast$Liangliang Yuan. Email: llyuan@shao.ac.cn\and
	\small$^\ast$Shuanggen Jin. Email: sgjin@hpu.edu.cn\and
}

\begin{document} 

\maketitle

\begin{abstract} \bfseries \boldmath
The ambipolar electrostatic field has long been recognized as a key driver of ion escape from planetary atmospheres. Elucidating the mechanisms responsible for the generation of this field is critical for understanding atmospheric escape and the evolution of habitability on terrestrial planets. Yet, existing comparisons between ambipolar diffusion theory and in-situ electric potential measurements have largely neglected the effect of electron heat flow. Confronting the theory incorporating heat-flow effect with in-situ electrical potential data from the \textit{Endurance} sounding rocket mission, we identify observational signatures of electron heat-flow effects. Furthermore, the implications of electron heat-flow effect across terrestrial planets are revealed, focusing on its capacity to resolve the enigma of Venusian electric potential drop anomaly. The anisotropic ion temperatures and the associated enhancement of electron heat-flow effect could potentially explain the anomalous electric potential drop observed in the ionosphere of Venus.
\end{abstract}

\noindent
The ambipolar electrostatic field of the Earth has long been considered as one of the primary drivers of ion escape, particularly along open geomagnetic field lines \cite{axford1968, Banks1968}. Although multiple attempts have been made to directly measure dc electric fields in Earth’s ionosphere \cite{mende1968experimental,fung1991search}, the first and thus far only comprehensive in-situ observations using atmospheric photoelectron spectroscopy were achieved recently \cite{collinson2022endurance}. As is reported in the initial results, the measured ambipolar electric field is shown to be one of the most important driving forces for the cold ion escape and appears to be aligned with predictions based on classical first-order ambipolar theory \cite{collinson2024Earth}.

However, in existing comparisons between ambipolar diffusion theory and in-situ potential measurements, the influence of electron heat flow on the ambipolar electric field has typically not been incorporated \cite{collinson2019,collinson2022endurance}. Owing to their low mass, electrons in the ionosphere exhibit substantial temperature variability, which can be significantly influenced by various heating processes, including auroral particle precipitation, photoelectron heating and thermal conduction along magnetic field lines \cite{RR1975, khazanov1997}. Electron heat flow is expected to be closely coupled with the plasma momentum equations, a coupling commonly referred to as the thermoelectric effect \cite{Conrad1979, schunk2000}. Nevertheless, direct observational signature of electron heat-flow effect on ambipolar electric field has yet to be detected, and quantitative comparisons between model predictions and measurements remain unavailable.

For small deviations from spatial uniformity and from Maxwellian velocity distributions, perturbation methods yield closed-form solutions of the Boltzmann equation \cite{Grad1949, burgers1969flow}. Here a system of diffusion equations for a partially ionized atmosphere is derived using the eight-moment approximation. Based on the proposed transport theory incorporating electron heat flow, we identify observational signatures of electron heat-flow effects on ambipolar electric fields through comparisons with in-situ electrical potential data from the \textit{Endurance} sounding rocket mission (see supplementary materials for details).

\section*{Observational Signatures}
For a $\mathrm{O^+}$- or $\mathrm{O_2^+}$-dominated ionosphere, a system of simplified diffusion equation is derived with heat-flow effect. The derivation of the ambipolar electric field with heat-flow effect is presented in supplementary materials and methods. Due to firings of the attitude control system (ACS) thruster, potential measurements from both the photoelectron spectrometers (PES) and the swept Langmuir probe (SLP) were interrupted multiple times during the flight. To avoid any uncertainties related to the data gaps, our analysis is restricted to the measurement segments of the \textit{Endurance} mission below 40 km. Therefore, no additional uncertainty due to the data gap contaminates the data–model comparisons.

\begin{figure} 
	\centering
	\includegraphics[width=0.8\textwidth]{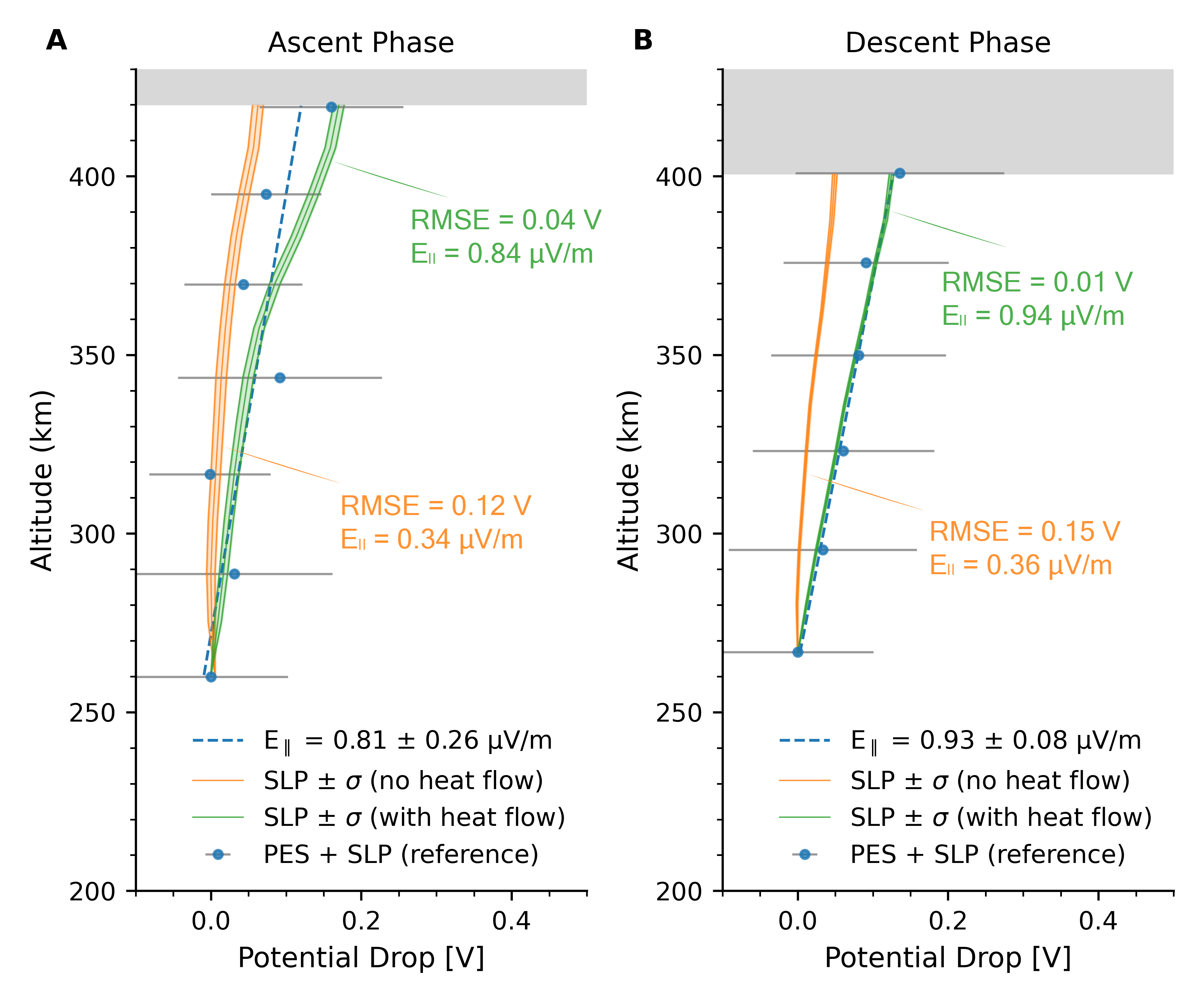} 

	\caption{\textbf{Electric potential drop profiles measured by PES and SLP onboard \textit{Endurance} and corresponding theoretical predictions.}
		 (A) Potential drop profiles during the ascent phase; (B) Potential drop profiles during the descent phase. The blue data points, accompanied by horizontal error bars, represent the electric potential drop values measured by PES and subsequently corrected using SLP. The blue dashed lines represent the linear regression lines based on PES measurements. And the orange and green regions represent the predictions of the ambipolar electrostatic potential drop based on field-aligned transport theory without and with heat-flow effects, respectively. The shaded regions represent the altitude range of ACS fire. The colored texts indicate the root mean square errors (RMSE) relative to the corresponding PES potential drop measurements, and mean values of predicted electric fields. The major ion is assumed to be $\mathrm{O^+}$. }
	\label{fig:2} 
\end{figure}

Figure \ref{fig:2}a,b shows the ionospheric electrostatic potential drop based on the PES measurements. The blue dots denote calibrated potential drop profiles derived from PES and SLP potential measurements, while the blue dashed lines indicate the corresponding linear regression. The potential drop measurements exhibited greater fluctuation during the ascent phase, with a regression slope of $0.81 \pm 0.26~\mu$V/m. In contrast, the measurements during the descent phase were smoother, corresponding to a linear regression slope of $0.93 \pm 0.08~\mu$V/m. The orange and green regions represent the predictions of the ambipolar electrostatic potential drop based on ambipolar theory without and with heat-flow effects, respectively. During both the ascent and descent phases of the mission, we observed that predictions derived from ambipolar theory without heat-flow effect (Fig.~\ref{fig:2}a,b, orange region) are consistently smaller than the PES measurements (Fig.~\ref{fig:2}a,b, blue dots), with the discrepancy being more pronounced during the descent phase (Fig.~\ref{fig:2}b). Such a discrepancy is also evident in the predicted electron density profiles (Fig.~\ref{fig:profile_endurance}, orange lines). These characteristics align with the initial results from the \textit{Endurance} mission \cite{collinson2024Earth}. Compared to the classical model neglecting heat-flow effects (Fig.~\ref{fig:2}a,b, orange region), predictions incorporating heat-flow effects (Fig.~\ref{fig:2}a,b, green region) exhibit improved agreement with the PES measurements. This improvement is also observable in the electron density profile (Fig.~\ref{fig:profile_endurance}, green lines). Incorporating electron heat flow eliminates the previous underestimation of the electric field during both the ascent and descent phases. The improvement of data-model consistency underscores the substantial role of electron heat flow in modulating the ambipolar electric fields in Earth’s ionosphere. Notably, we emphasize that collisions with neutral particles play an essential role in modulating the ambipolar electric field. In a fully ionized ionosphere where both ion-neutral and electron-neutral collisions are absent (Fig.~\ref{fig:S4}), the electron heat-flow effect is reduced by 33\% (ascent phase) and 37\% (descent phase).

\section*{Modulation Criterion}
According to ambipolar diffusion theory, in an isothermal hydrostatic ionosphere composed of a single ion species, electrons and ions exhibit identical scale heights.

\begin{figure} 
	\centering
	\includegraphics[width=0.9\textwidth]{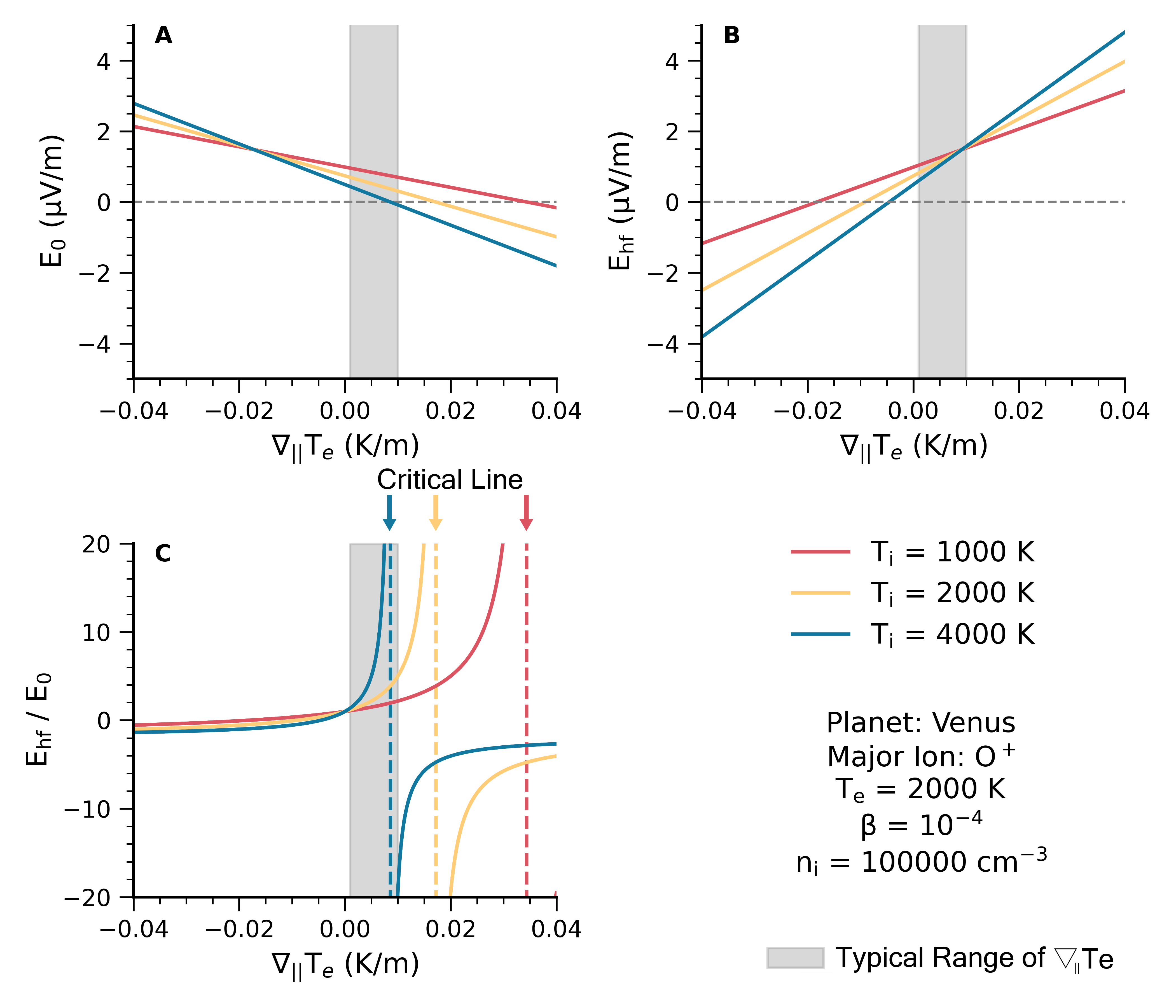} 

	\caption{\textbf{Predicted ambipolar electric fields with ($\mathrm{E_{hf}}$) and without ($\mathrm{E_0}$) heat-flow effect as a function of electron temperature gradient (\(\nabla_{\parallel} \mathrm{T_e}\)) and ion temperature $\mathrm{T_i}$.} 
	(A) Ambipolar electric fields without heat-flow effect ($\mathrm{E_0}$); 
	(B) Ambipolar electric fields with heat-flow effect ($\mathrm{E_{hf}}$); 
	(C) Ratios of two electric fields (\(\mathrm{E_{hf}/E_0}\)).
The dashed vertical lines in panel C denote the critical lines. The major ion is assumed to be $\text{O}^+$ and electron temperature to be 2000~K in this simulation. Results are similar for $\mathrm{T_e} =$ 1000, 2000, and 4000~K.	The results are insensitive to the choice of the ionization degree ($\beta=\text{10}^{-4}$), as demonstrated in Fig.~\ref{fig:beta_simulation_O+} and \ref{fig:beta_simulation_O2+}.}
	\label{fig:3} 
\end{figure}

To illustrate the role of electron heat flow, the predicted electric fields as functions of ion temperature and electron temperature gradients are shown in Fig.~\ref{fig:3}. 
Both the electric field predictions incorporating heat-flow effects (Fig.~\ref{fig:3}b) and those neglecting them (Fig.~\ref{fig:3}a) exhibit a linear dependence on the electron temperature gradient. A notable feature is the emergence of a critical line (Fig.~\ref{fig:3}c, vertical dashed line) along which the ambipolar electric field predicted without heat-flow effects ($\mathrm{E_0}$, Fig.~\ref{fig:3}a) approaches zero.  According to Eq.~, smaller values of $\mathrm{T_e/T_i}$ correspond to a stronger heat-flow effect, while larger values indicate a weaker influence. This implies that in planetary ionospheres where electrons are colder than ions, the effect of heat flow is expected to be more pronounced compared to those where electrons are substantially hotter. On terrestrial planets such as Earth and Venus, the characteristic ionospheric electron temperature gradient is positive and on the order of $10^{-3}$ or $10^{-2}$~K/m (Fig.~\ref{fig:3}, shaded region). When the ion temperature is lower than the electron temperature, the critical line lies outside the typical range (Fig.~\ref{fig:3}c, shaded region), and no significant enhancement of the heat-flow effect occurs (Tabs.~\ref{tab:S1} to \ref{tab:S9}, $\mathrm{T_i/T_e}=0.5$). Similarly, when the electron temperature gradient is small (0.001~K/m), the electron heat-flow effect is generally weak, i.e., $\text{E}_\text{hf}$-to-$\text{E}_0$ ratio smaller than 1.5 (Tabs.~\ref{tab:S1}). The critical line falls within the typical range of electron temperature gradients only when the ion temperature exceeds 4000~K (Fig.~\ref{fig:3}c, blue dashed lines), twice the electron temperature. In this case, the predicted ambipolar electric field obtained without considering heat-flow effects (Fig.~\ref{fig:3}a) is significantly weaker than that derived when heat-flow effects are included (Fig.~\ref{fig:3}b). Near the critical line, the primary effect of electron heat flow is to enhance the upward ambipolar electric field while maintaining overall momentum conservation of the plasma.

\section*{Venusian Electric Potential Drop Anomaly}
Recent studies indicate that the transient electric potential drop observed in the Venusian ionosphere exceeds the predicted value of 0.9~V ($\sim$0.3~$\mu$V/m) from classical ambipolar electric field theory by more than a factor of five \cite{collinson2019, collinson2023}. And the majority of the strong electric field measurements occurred near the solar terminator. Currently, no physically plausible explanation exists for this significant discrepancy between observations and model predictions. 
We propose that the unexplained electric potential drop anomaly observed in the ionosphere of Venus likely results from enhanced electron heat-flow effects. On one hand, the dense neutral atmosphere of Venus facilitates rapid electron cooling through frequent collisions with neutrals, leading to a steep electron temperature gradient ($\sim$0.04~K/m) below 200~km \cite{knudsen1979venus}. This rapid cooling may broaden the permissible range of $\nabla_{\parallel} \mathrm{T_e}$ toward higher values and amplify the heat-flow modulation effect, which suggests that the electron heat-flow effect is confined to altitudes below approximately 200 km, where enhanced electron cooling becomes significant (see Fig.~\ref{fig:schematic}). This argument explains why the transient electric field exhibits altitude independence and is detectable at low altitudes below approximately 200~km. On the other hand, the relative ion-neutral flow, e.g., supersonic ion flows near the solar terminator \cite{knudsen1980venus}, and associated ion temperature anisotropy can produce a plasma environment that enlarges the $\mathrm{T_i/T_e}$ ratio at lower altitudes (below 220~km) \cite{schunk1981venus}. When the ion-neutral relative velocity aligns with the interplanetary magnetic field (IMF) direction, the parallel ion temperature may exceed the values measured by the retarding potential analyzer (RPA) by a factor of several, depending on the magnitude of the relative flow. This ion temperature enhancement enables the ambipolar modulation criterion to be met. The alignment requirement imposed by this argument partly explains the low occurrence rate ($\sim$1\%) of strong electric fields \cite{collinson2023}. A potential means of validating this hypothesis is through the simultaneous measurement of magnetic field, plasma temperature anisotropy, and ion drift velocity.

Quantitatively, when the electron temperature gradient reaches 0.01~K/m (see Tabs.~\ref{tab:S7} to~\ref{tab:S9}), the electric field without heat-flow effect ($\text{E}_0$) is smaller than that with heat-flow effect ($\text{E}_{\text{hf}}$) by a factor of five for Venus under the condition $\mathrm{T_i/T_e}=1$. The predicted ambipolar electric field is 0.31~$\mu$V/m without the heat-flow effect and 1.55~$\mu$V/m when the heat-flow effect is included. These values are quantitatively consistent with both the peak electric fields derived from classical ambipolar theory and plasma measurements on the \textit{Pioneer Venus Orbiter} ($\sim$0.3~$\mu$V/m), and those required to account for the transient electric potential drops detected by the \textit{Venus Express} spacecraft ($\sim$1.5~$\mu$V/m) \cite{collinson2019}.




\clearpage 

%
\bibliography{science_template} 
\bibliographystyle{sciencemag}

%
%
%
%
%
%


\section*{Acknowledgments}
The authors would like to thank NASA Goddard Space Flight Center (GSFC) for providing the \textit{Endurance} data. We also thank Dr. Glyn Collinsion for useful discussions on the data interpretation.
\paragraph*{Funding:}
This research was supported by the National Natural Science Foundation of China (NSFC) (Grant No. 42504171), the Natural Science Foundation of Shanghai (Grant No. 24YF2754300, Grant No. 24PJA152). L.Y. also acknowledges support from the Pioneering Project of the Chinese Academy of Sciences.
\paragraph*{Author contributions:}
The conception was conceived by L. Y., who acquired the funding and drafted this paper. Data analysis was performed by L. Y. L. Y. and S. J.contributed to the review and editing works.
\paragraph*{Competing interests:}
There are no competing interests to declare.
\paragraph*{Data and materials availability:}
Endurance data used in this article are available at the Space Physics Data Facility of NASA through the Coordinated Data Analysis Web (CDAWeb) tool (https://cdaweb.gsfc.nasa.gov/).


\subsection*{Supplementary materials}
Materials and Methods\\
Figs. S1 to S11\\
Tabs. S1 to S9

\newpage


\renewcommand{\thefigure}{S\arabic{figure}}
\renewcommand{\thetable}{S\arabic{table}}
\renewcommand{\theequation}{S\arabic{equation}}
\renewcommand{\thepage}{S\arabic{page}}
\setcounter{figure}{0}
\setcounter{table}{0}
\setcounter{equation}{0}
\setcounter{page}{1} 


\begin{center}
\section*{Supplementary Materials for\\ \scititle}

Liangliang Yuan$^{\ast}$,
Shuanggen Jin$^{\ast}$\\ 
\small$^\ast$Liangliang Yuan. Email: llyuan@shao.ac.cn\\
\small$^\ast$Shuanggen Jin. Email: sgjin@hpu.edu.cn\\
\end{center}

\subsubsection*{This PDF file includes:}
Materials and Methods\\
Figs. S1 to S11\\
Tabs. S1 to S9
\newpage

\begin{figure} 
	\centering
	\includegraphics[width=0.9\textwidth]{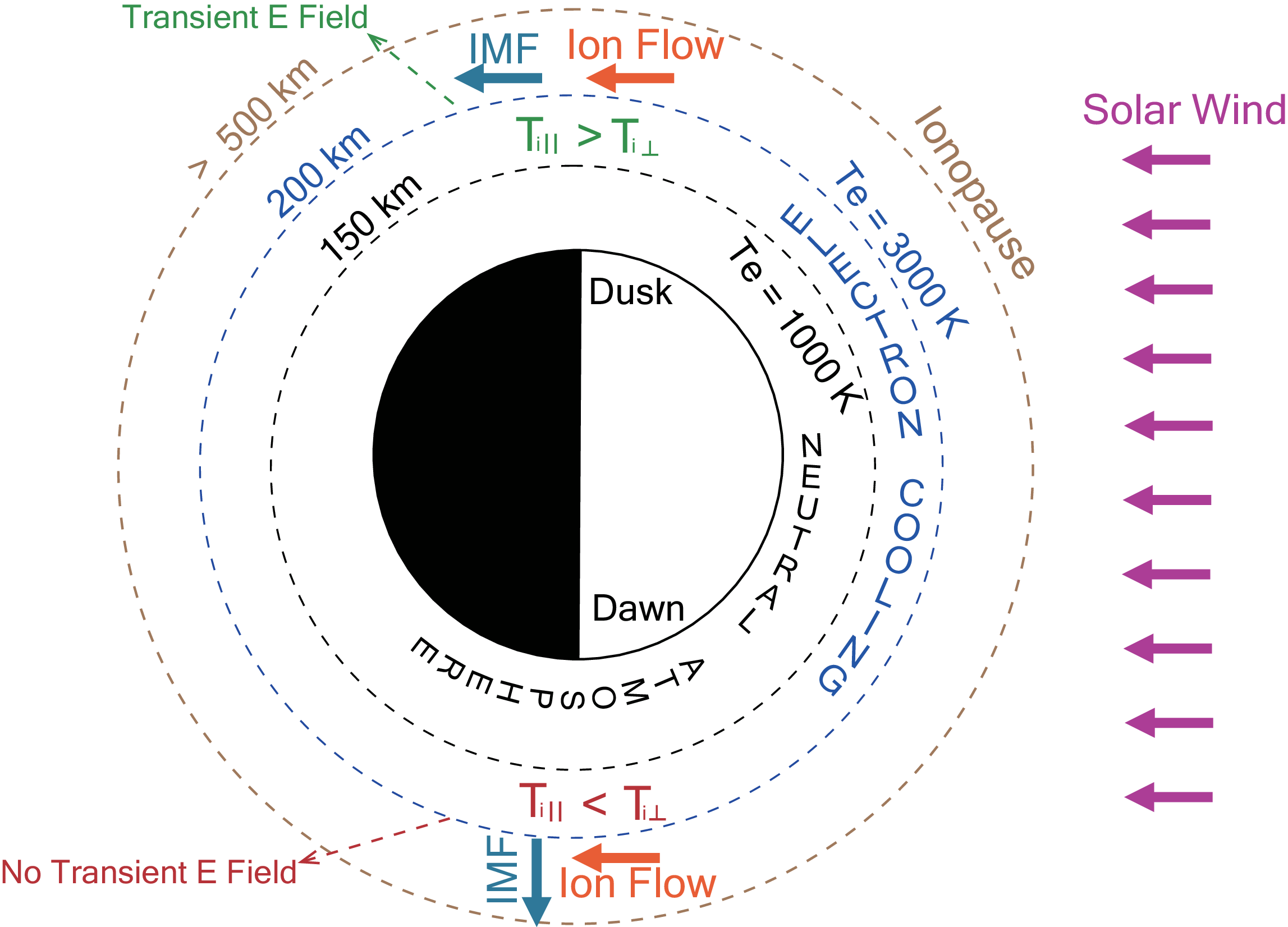} 
\caption{\textbf{Schematic explanation of the Venusian electric potential drop anomaly.} 
    The illustration summarizes the proposed mechanism, highlighting the key role of the enhancement of electron heat-flow effect in producing the transient electric field features. When the IMF aligns with the direction of the ion-neutral relative flow, the likelihood of a transient electric field generation is significantly greater than under the opposite condition.}
	\label{fig:schematic} 
\end{figure}

\begin{figure} 
	\centering
	\includegraphics[width=0.7\textwidth]{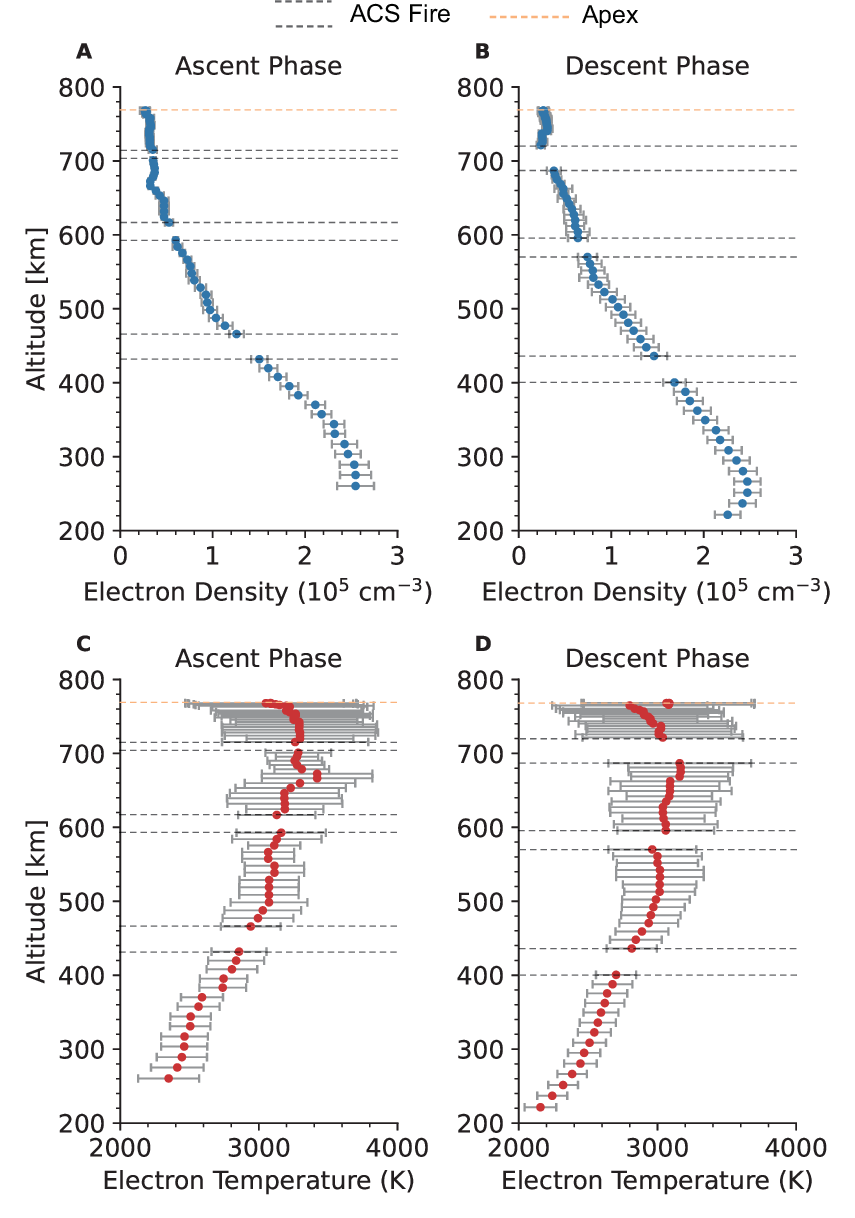} 
	\caption{\textbf{Electron density and electron temperature profiles measured by swept Langmuir probe onboard \textit{Endurance}.}
		 (A) Electron density profile in ascent phase of the rocket; (B) Electron density profile in descent phase of the rocket; (C) Electron temperature profile in ascent phase of the rocket; (D) Electron temperature profile in descent phase of the rocket. The black dashed lines indicate the altitude ranges affected by ACS thruster firings, while the yellow dashed line marks the trajectory apex of the \textit{Endurance} sounding rocket.}
	\label{fig:S1} 
\end{figure}

\begin{figure} 
	\centering
	\includegraphics[width=0.8\textwidth]{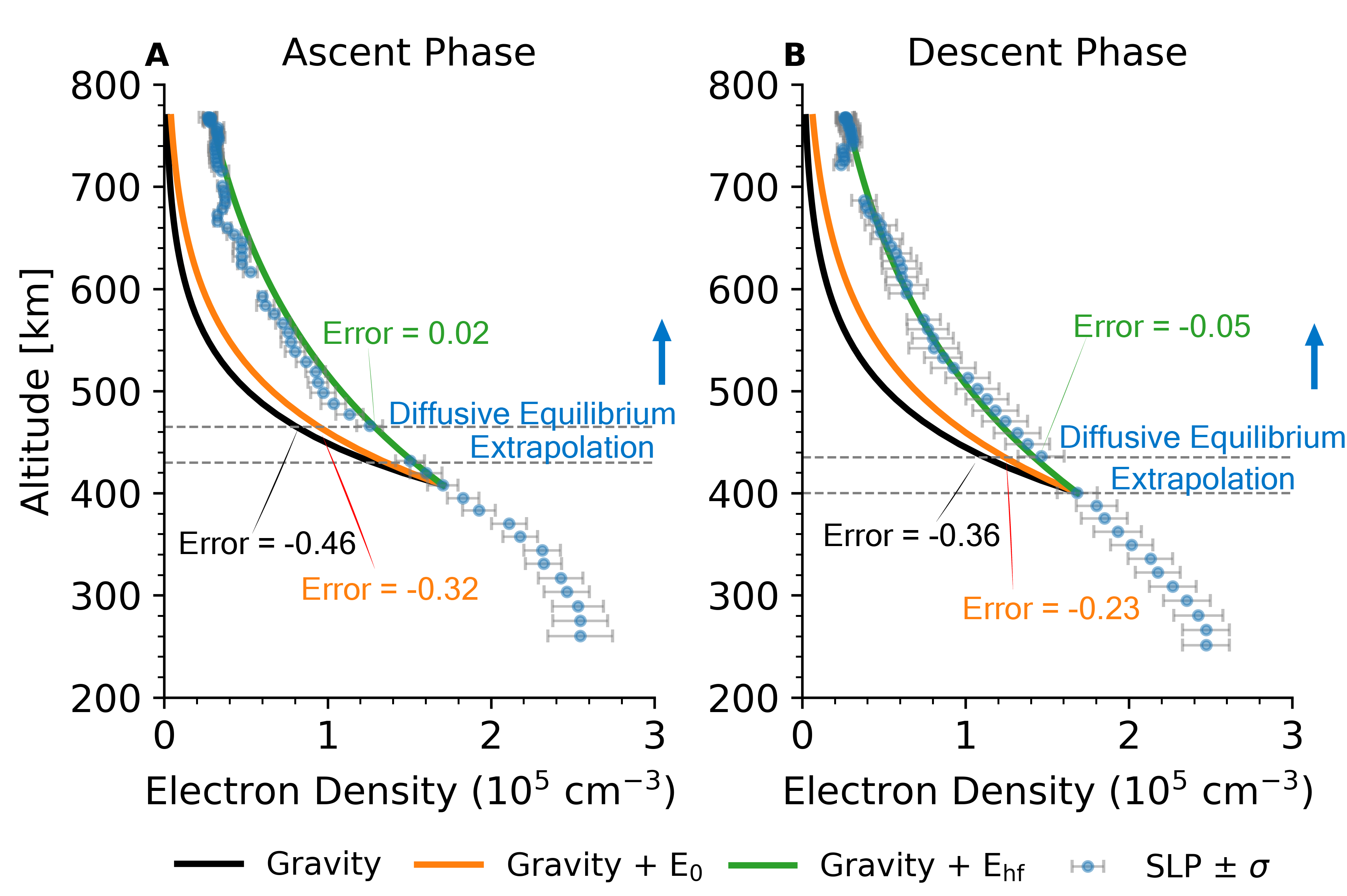} 
	\caption{\textbf{Measured (blue dots) and theoretically predicted (using Eq.~, black line: gravity only; orange line: gravity and electric field without heat-flow effect, green line: gravity and electric field with heat-flow effect) electron density profiles along the \textit{Endurance} trajectory.}
    (A) Predicted electron density profiles in ascent phase of the rocket with $\mathrm{T_i}=1453$~K; (B) Predicted electron density profiles in descent phase of the rocket with $\mathrm{T_i}=1614$~K; The ion temperature range is derived from co-located EISCAT radar measurements, yielding values of $1453 \pm 376\text{~K}$ for the ascent phase and $1614 \pm 648\text{~K}$ for the descent phase \cite{collinson2024Earth}. The ionospheric plasma is assumed to be in diffusive equilibrium above 400~km whereas photo-ionization dominates near the peak density height. Between the dashed lines is the extrapolation region. The colored numerals indicate the respective extrapolation error at the upper dashed line.}
	\label{fig:profile_endurance} 
\end{figure}

\begin{figure} 
	\centering
	\includegraphics[width=0.8\textwidth]{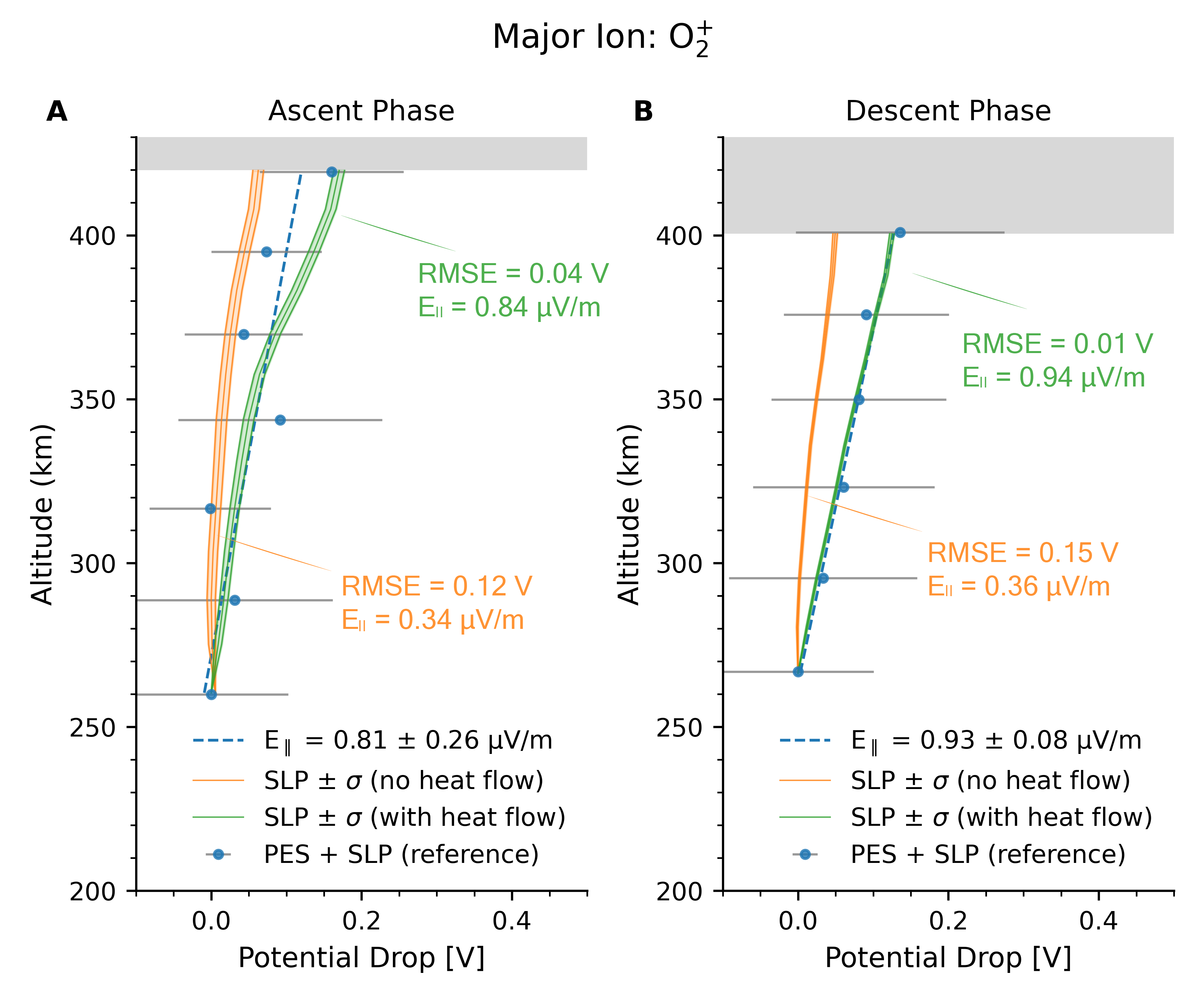} 

	\caption{\textbf{Electric potential drop profiles measured by PES onboard \textit{Endurance} and corresponding theoretical estimates, but for $\mathrm{O_2^+}$-dominated ionosphere.}
		 (A) Potential drop profiles during the ascent phase; (B) Potential drop profiles during the descent phase. The blue dashed lines represent the linear regression lines based on PES measurements. And the orange and green regions represent the predictions of the ambipolar electrostatic potential drop based on field-aligned transport theory without and with heat-flow effects, respectively. The shaded region represents the altitude range of ACS fire.}
	\label{fig:S3} 
\end{figure}

\begin{figure} 
	\centering
	\includegraphics[width=0.8\textwidth]{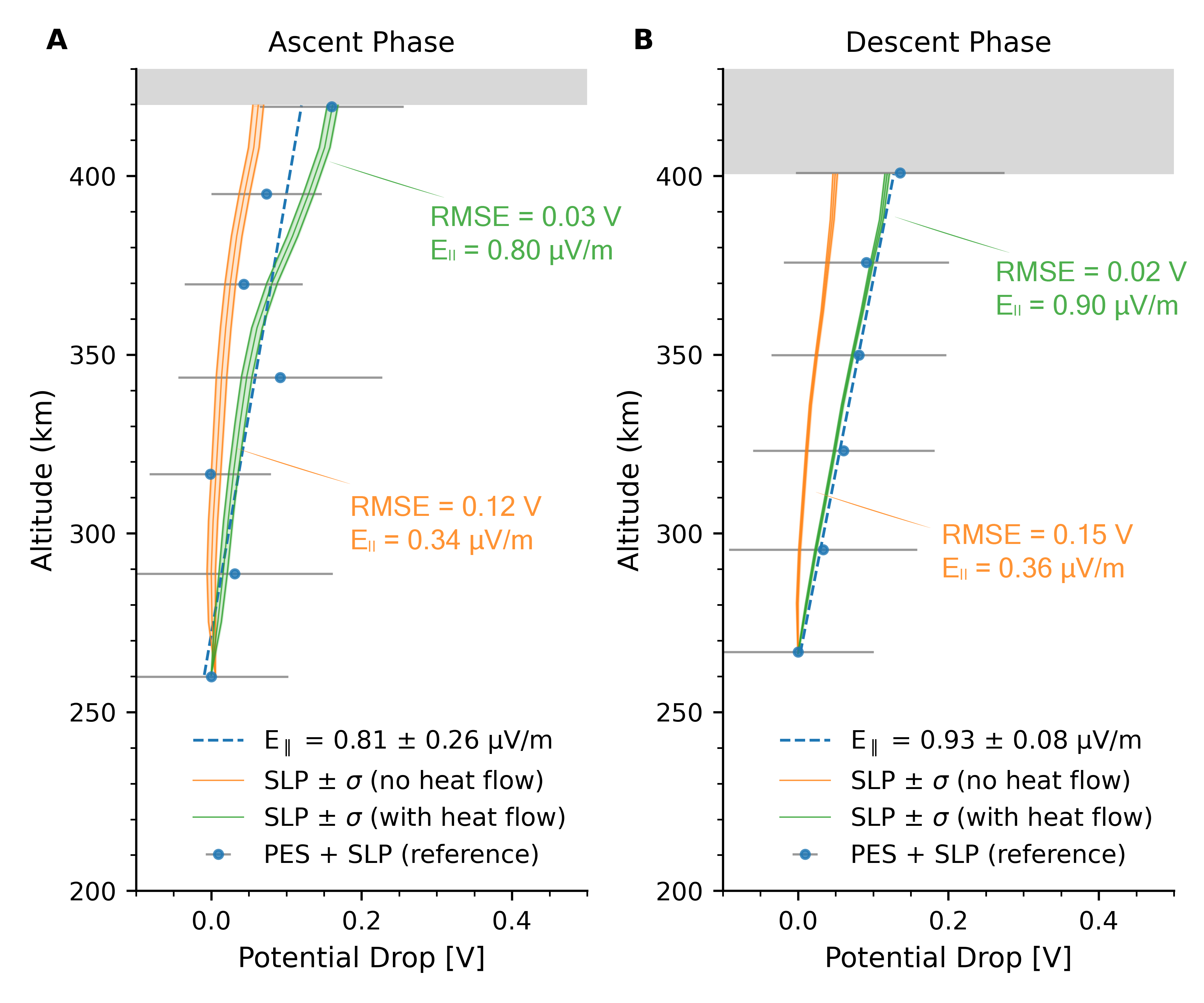} 

	\caption{\textbf{Electric potential drop profiles measured by PES onboard \textit{Endurance} and corresponding theoretical estimates, but for $\eta=10$.}
		 (A) Potential drop profiles during the ascent phase; (B) Potential drop profiles during the descent phase. The blue dashed lines represent the linear regression lines based on PES measurements. And the orange and green regions represent the predictions of the ambipolar electrostatic potential drop based on field-aligned transport theory without and with heat-flow effects, respectively. The shaded regions represent the altitude range of ACS fire. The colored texts indicate the root mean square errors (RMSE) relative to the corresponding PES potential drop measurements, and mean values of predicted electric fields. }
	\label{fig:eta_10} 
\end{figure}

\begin{figure} 
	\centering
	\includegraphics[width=0.8\textwidth]{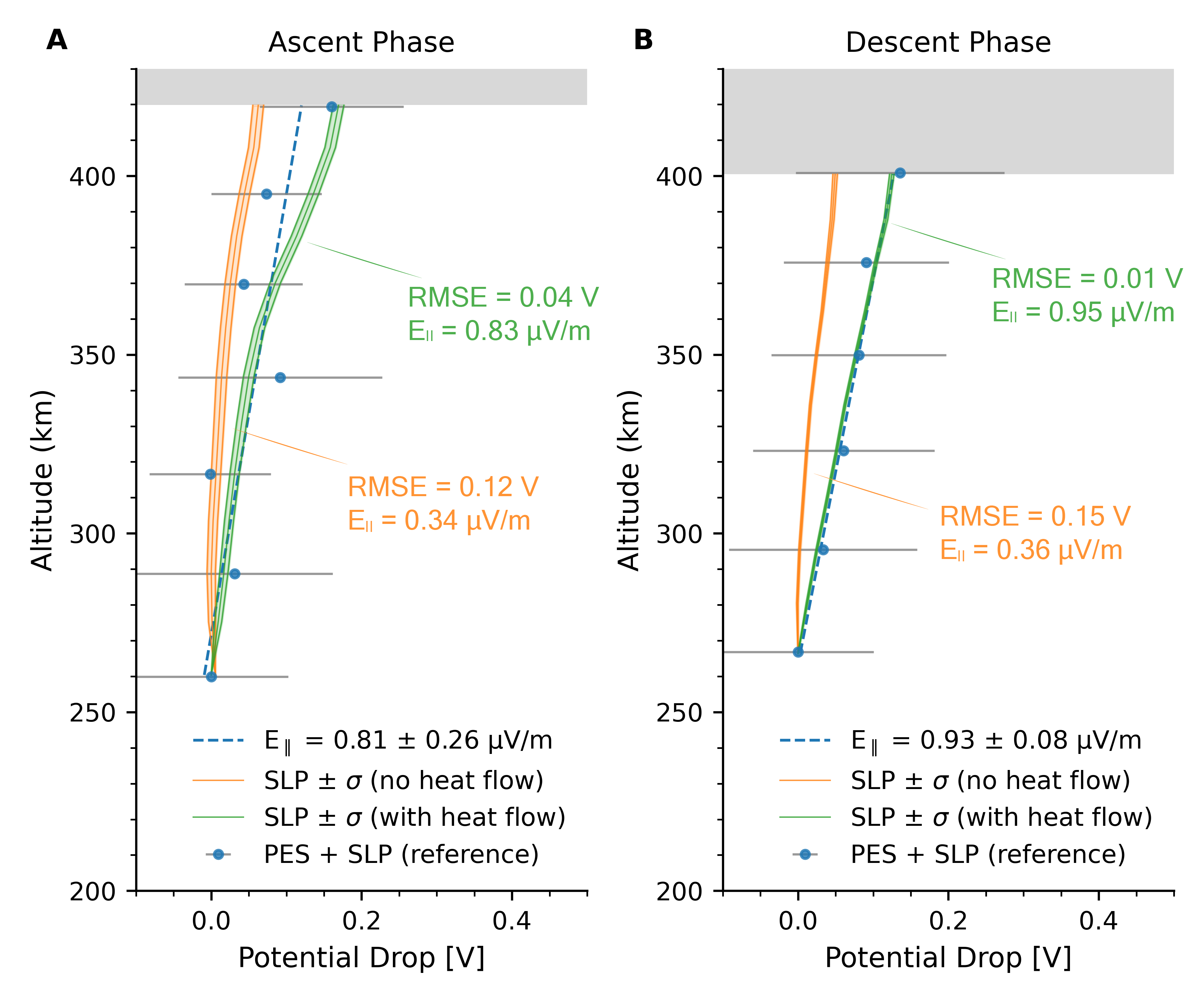} 

	\caption{\textbf{Electric potential drop profiles measured by PES onboard \textit{Endurance} and corresponding theoretical estimates, but for $\eta=0.1$.}
		 (A) Potential drop profiles during the ascent phase; (B) Potential drop profiles during the descent phase. The blue dashed lines represent the linear regression lines based on PES measurements. And the orange and green regions represent the predictions of the ambipolar electrostatic potential drop based on field-aligned transport theory without and with heat-flow effects, respectively. The shaded regions represent the altitude range of ACS fire. The colored texts indicate the root mean square errors (RMSE) relative to the corresponding PES potential drop measurements, and mean values of predicted electric fields. }
	\label{fig:eta_0.1} 
\end{figure}

\begin{figure} 
	\centering
	\includegraphics[width=0.8\textwidth]{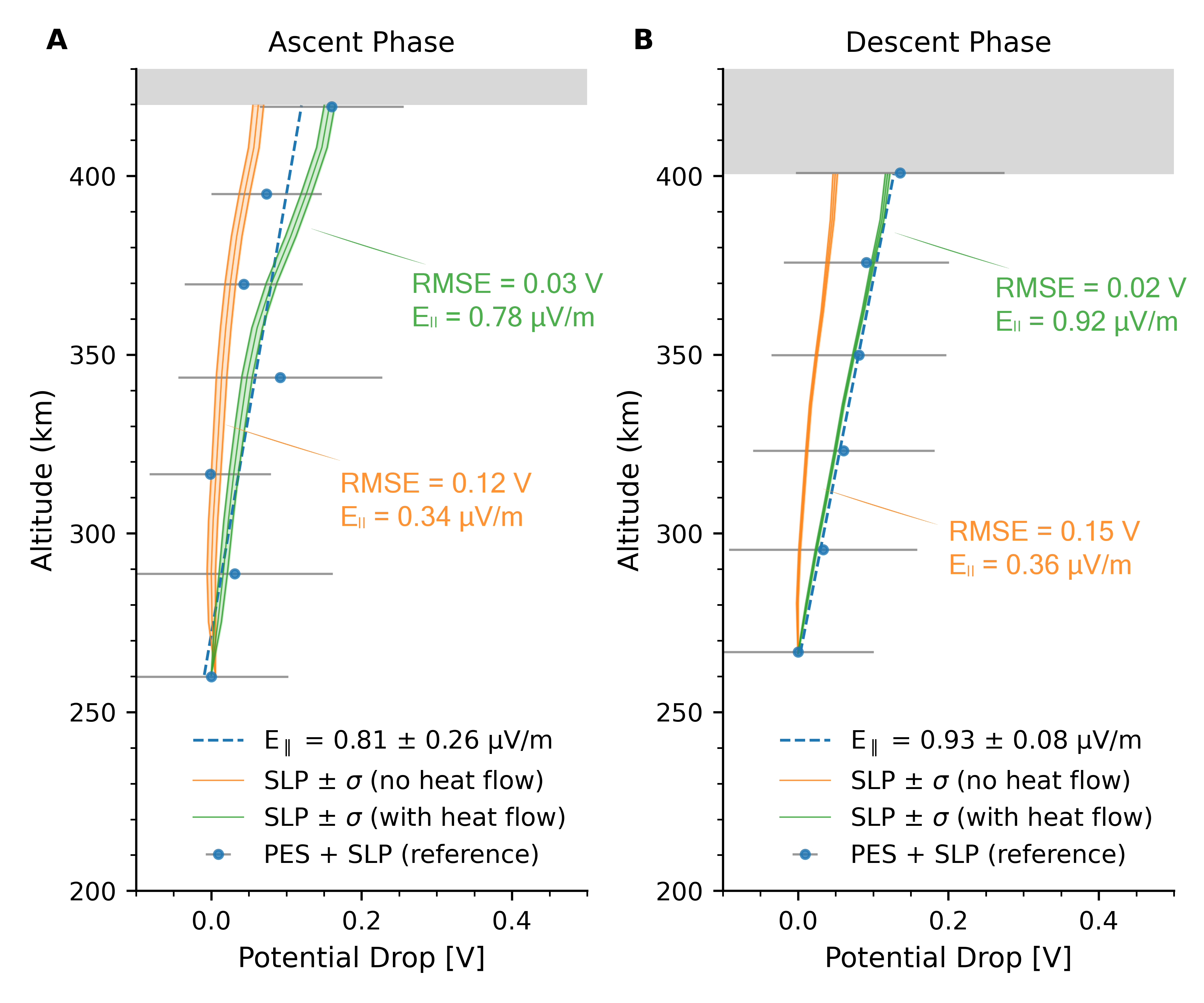} 

	\caption{\textbf{Electric potential drop profiles measured by PES onboard \textit{Endurance} and corresponding theoretical estimates, but for $\eta=0.01$.}
		 (A) Potential drop profiles during the ascent phase; (B) Potential drop profiles during the descent phase. The blue dashed lines represent the linear regression lines based on PES measurements. And the orange and green regions represent the predictions of the ambipolar electrostatic potential drop based on field-aligned transport theory without and with heat-flow effects, respectively. The shaded regions represent the altitude range of ACS fire. The colored texts indicate the root mean square errors (RMSE) relative to the corresponding PES potential drop measurements, and mean values of predicted electric fields. }
	\label{fig:eta_0.01} 
\end{figure}

\begin{figure} 
	\centering
	\includegraphics[width=0.5\textwidth]{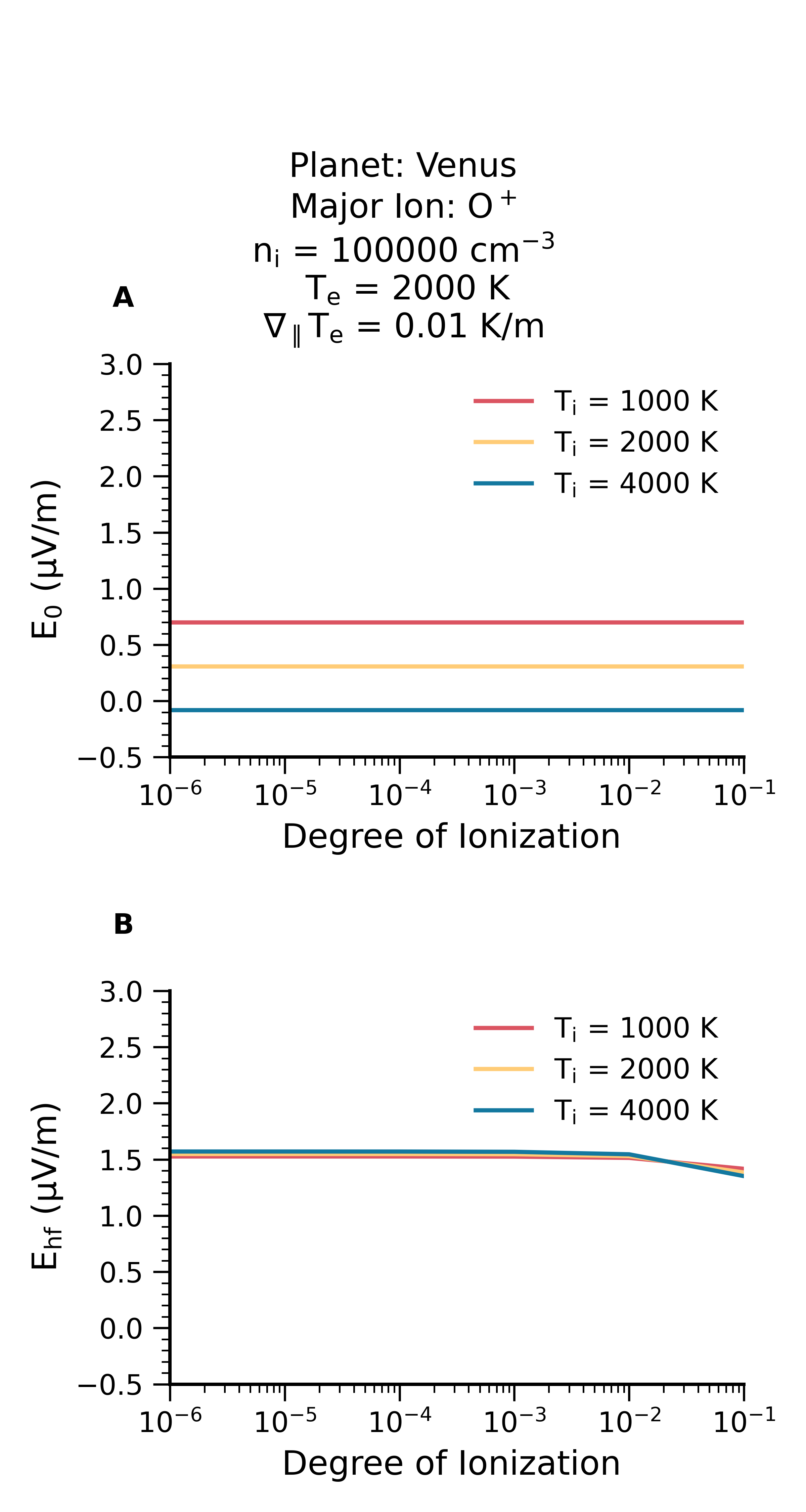} 

	\caption{\textbf{Predicted ambipolar electric fields with ($\mathrm{E_{hf}}$) and without ($\mathrm{E_0}$) heat-flow effect as a function of degree of ionization and ion temperature $\mathrm{T_i}$ on Venus.} 
	(A) Ambipolar electric fields without heat-flow effect ($\mathrm{E_0}$); 
	(B) Ambipolar electric fields with heat-flow effect ($\mathrm{E_{hf}}$); 
The major ion is assumed to be $\text{O}^+$ and electron temperature to be 2000~K in this simulation. Results are similar for $\mathrm{T_e} =$ 1000, 2000, and 4000~K.}
	\label{fig:beta_simulation_O+} 
\end{figure}

\begin{figure} 
	\centering
	\includegraphics[width=0.5\textwidth]{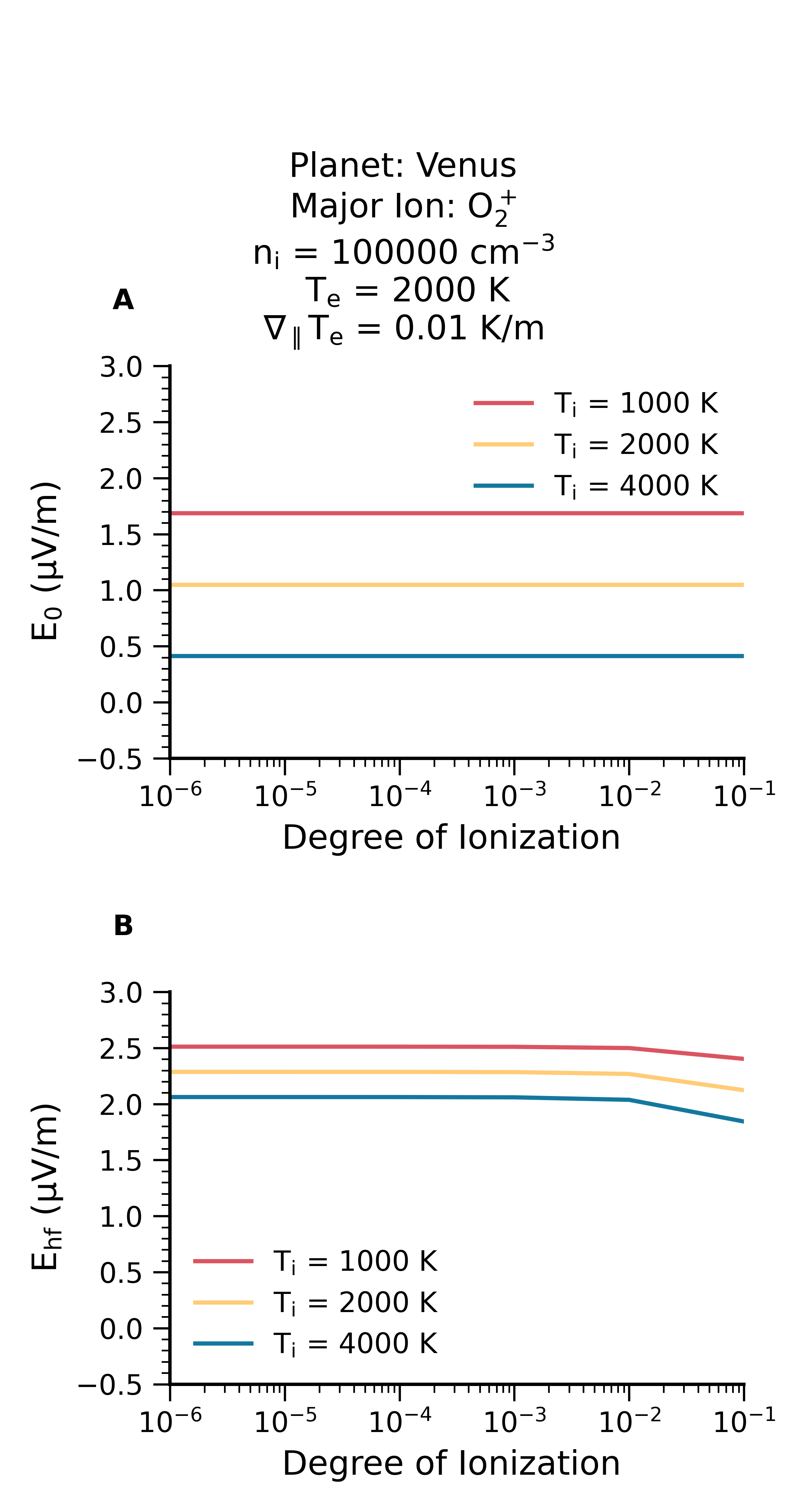} 

	\caption{\textbf{Predicted ambipolar electric fields with ($\mathrm{E_{hf}}$) and without ($\mathrm{E_0}$) heat-flow effect as a function of degree of ionization and ion temperature $\mathrm{T_i}$ on Venus.} 
	(A) Ambipolar electric fields without heat-flow effect ($\mathrm{E_0}$); 
	(B) Ambipolar electric fields with heat-flow effect ($\mathrm{E_{hf}}$); 
The major ion is assumed to be $\text{O}_2^+$ and electron temperature to be 2000~K in this simulation. Results are similar for $\mathrm{T_e} =$ 1000, 2000, and 4000~K.}
	\label{fig:beta_simulation_O2+} 
\end{figure}

\begin{figure} 
	\centering
	\includegraphics[width=0.8\textwidth]{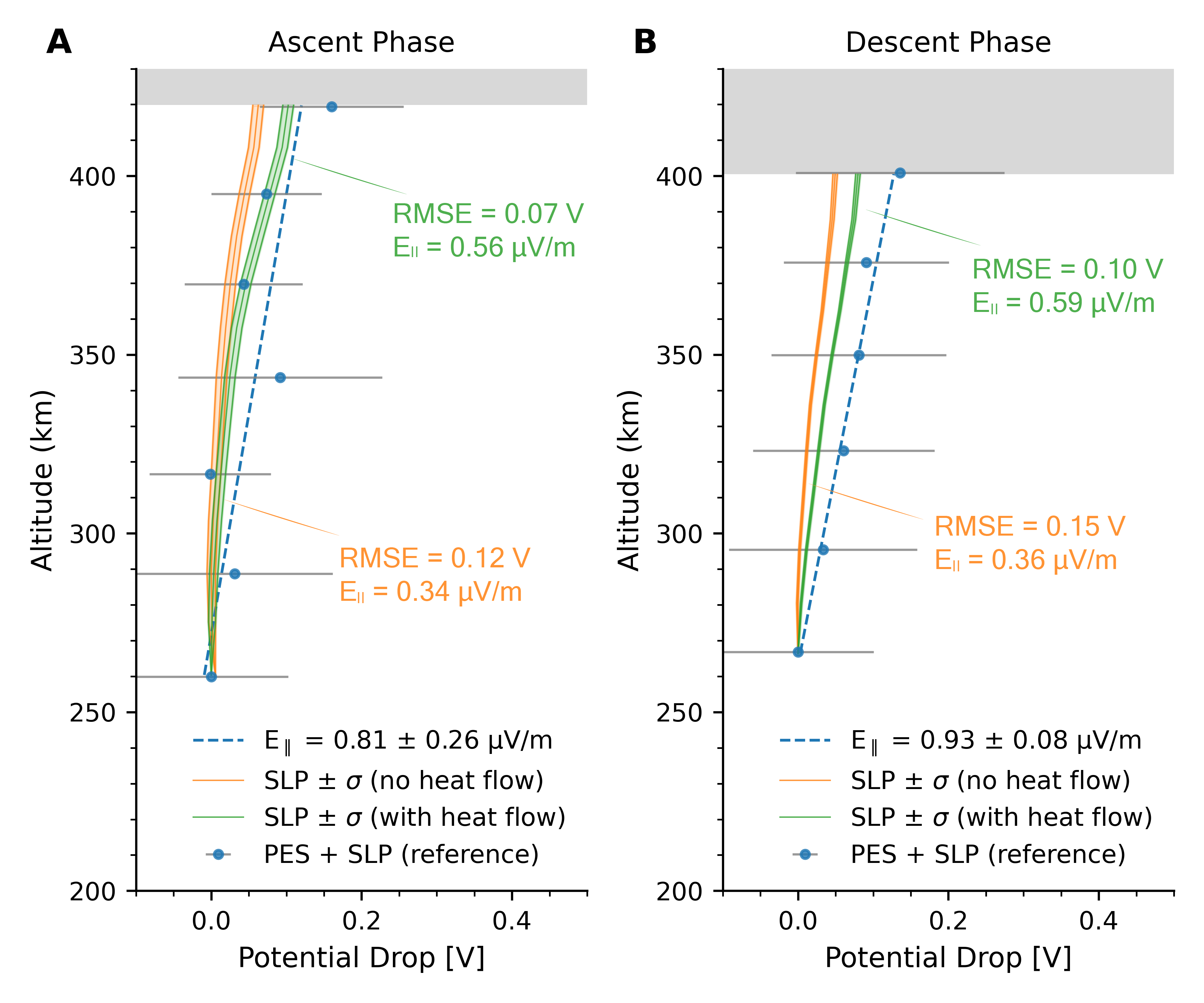} 

	\caption{\textbf{Electric potential drop profiles measured by PES onboard \textit{Endurance} and corresponding theoretical estimates, but for fully-ionized ionosphere ($\beta=1$).}
		 (A) Potential drop profiles during the ascent phase; (B) Potential drop profiles during the descent phase. The blue dashed lines represent the linear regression lines based on PES measurements. And the orange and green regions represent the predictions of the ambipolar electrostatic potential drop based on field-aligned transport theory without and with heat-flow effects, respectively. The shaded regions represent the altitude range of ACS fire. The colored texts indicate the root mean square errors (RMSE) relative to the corresponding PES potential drop measurements, and mean values of predicted electric fields. }
	\label{fig:S4} 
\end{figure}

\begin{figure} 
	\centering
	\includegraphics[width=1\textwidth]{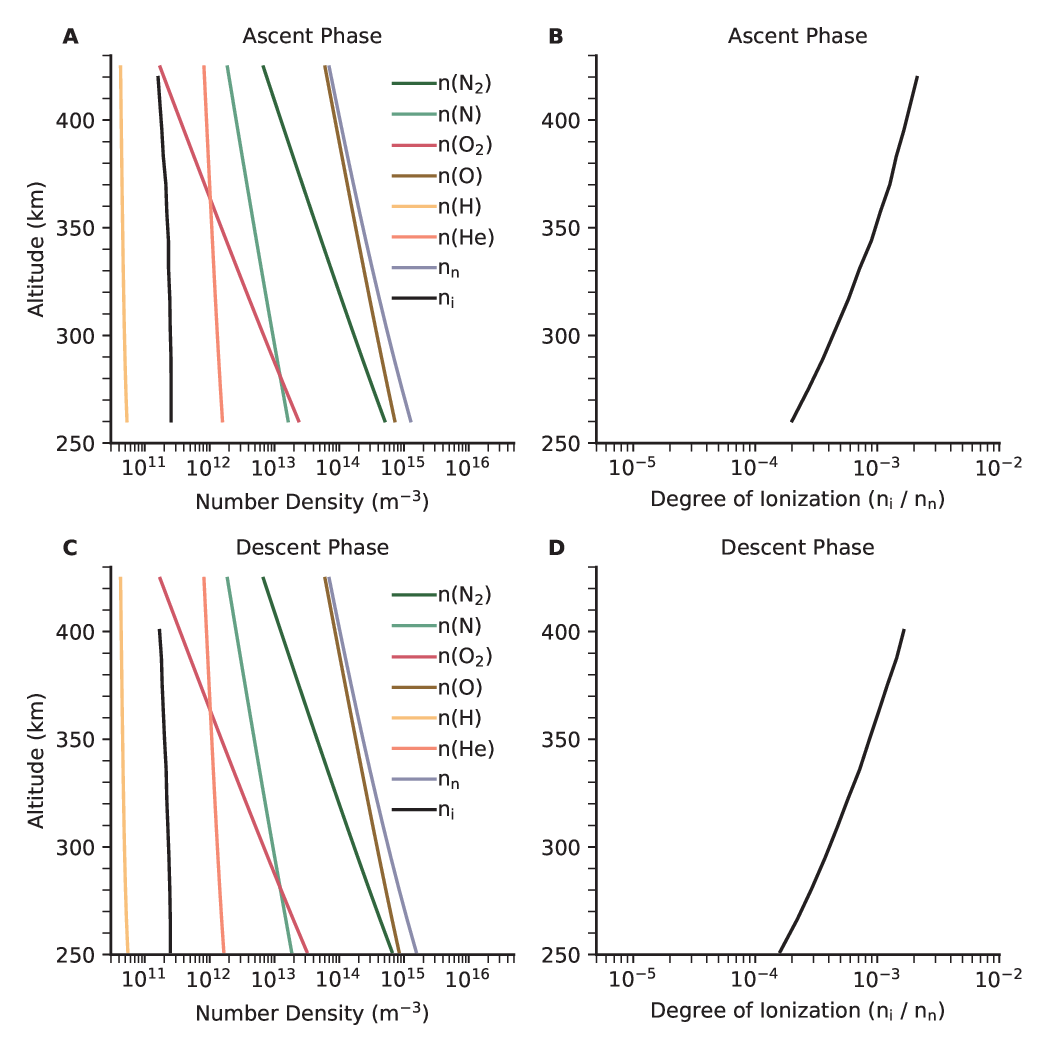} 
	\caption{\textbf{Number densities of major atmospheric species and corresponding degree of ionization along the \textit{Endurance} flight trajectory, derived from the NRLMSISE-00 model.}
		(A) Species number densities during ascent phase; (B) Degree of ionization during ascent phase; (C) Species number densities during descent phase; (D) Degree of ionization during descent phase.}
	\label{fig:S2} 
\end{figure}

\begin{table}
\centering
\caption{$E_{\text{hf}}$-to-$E_0$ ratio at a fixed ionization degree ($\beta=10^{-4}$), electron temperature ($T_e=2000$~K) and electron temperature gradient ($\nabla T_e = 0.001~\text{K/m}$) for different planetary ionospheres and ion species under varying ion-to-electron temperature ratios.}
\renewcommand{\arraystretch}{1.2}
\begin{tabular}{lcccr}
\\
\hline
Planet & Ion &  $\mathrm{T_i/T_e}=0.5$ & $\mathrm{T_i/T_e}=1$ & $\mathrm{T_i/T_e}=2$ \\
\hline
Venus & O$_2^+$  & 1.0496 & 1.1021 & 1.2138 \\
Venus & O$^+$    & 1.1032 & 1.2151 & 1.4675 \\
Earth & O$_2^+$  & 1.0446 & 1.0917 & 1.1916 \\
Earth & O$^+$    & 1.0928 & 1.1928 & 1.4157 \\
\hline
\label{tab:S1}
\end{tabular}
\end{table}

\begin{table}
\centering
\caption{Values of $E_0$ at a fixed ionization degree ($\beta=10^{-4}$), electron temperature ($T_e=2000$~K) and electron temperature gradient ($\nabla T_e = 0.001~\text{K/m}$) for different planetary ionospheres and ion species under varying ion-to-electron temperature ratios. Units are in V/m.}
\renewcommand{\arraystretch}{1.2}
\begin{tabular}{lcccr}
\\
\hline
Planet & Ion &  $\mathrm{T_i/T_e}=0.5$ & $\mathrm{T_i/T_e}=1$ & $\mathrm{T_i/T_e}=2$ \\
\hline
Venus & O$_2^+$ & 1.9407e-06 & 1.4294e-06 & 9.1810e-07 \\
Venus & O$^+$    & 9.5292e-07 & 6.8858e-07 & 4.2423e-07 \\
Earth & O$_2^+$ & 2.1500e-06 & 1.5864e-06 & 1.0228e-06 \\
Earth & O$^+$    & 1.0576e-06 & 7.6708e-07 & 4.7657e-07 \\
\hline
\label{tab:S2}
\end{tabular}
\end{table}

\begin{table}
\centering
\caption{Values of $E_{hf}$ at a fixed ionization degree ($\beta=10^{-4}$), electron temperature ($T_e=2000$~K) and electron temperature gradient ($\nabla T_e = 0.001~\text{K/m}$) for different planetary ionospheres and ion species under varying ion-to-electron temperature ratios. Units are in V/m.}
\renewcommand{\arraystretch}{1.2}
\begin{tabular}{lcccr}
\\
\hline
Planet & Ion &  $\mathrm{T_i/T_e}=0.5$ & $\mathrm{T_i/T_e}=1$ & $\mathrm{T_i/T_e}=2$ \\
\hline
Venus & O$_2^+$  & 2.0370e-06 & 1.5753e-06 & 1.1145e-06 \\
Venus & O$^+$    & 1.0512e-06 & 8.3668e-07 & 6.2257e-07 \\
Earth & O$_2^+$ & 2.2459e-06 & 1.7317e-06 & 1.2187e-06 \\
Earth & O$^+$    & 1.1557e-06 & 9.1495e-07 & 6.7470e-07 \\
\hline
\label{tab:S3}
\end{tabular}
\end{table}

\begin{table}
\centering
\caption{$E_{\text{hf}}$-to-$E_0$ ratio at a fixed ionization degree ($\beta=10^{-4}$), electron temperature ($T_e=2000$~K) and electron temperature gradient ($\nabla T_e = 0.005~\text{K/m}$) for different planetary ionospheres and ion species under varying ion-to-electron temperature ratios.}
\renewcommand{\arraystretch}{1.2}
\begin{tabular}{lcccr}
\\
\hline
Planet & Ion &  $\mathrm{T_i/T_e}=0.5$ & $\mathrm{T_i/T_e}=1$ & $\mathrm{T_i/T_e}=2$ \\
\hline
Venus & O$_2^+$  & 1.2360 & 1.5154 & 2.2596 \\
Venus & O$^+$    & 1.5169 & 2.2619 & 5.5077 \\
Earth & O$_2^+$  & 1.2115 & 1.4578 & 2.0923 \\
Earth & O$^+$    & 1.4592 & 2.0944 & 4.3930 \\
\hline
\label{tab:S4}
\end{tabular}
\end{table}

\begin{table}[htbp]
\centering
\caption{Values of $E_0$ at a fixed ionization degree ($\beta=10^{-4}$), electron temperature ($T_e=2000$~K) and electron temperature gradient ($\nabla T_e = 0.005~\text{K/m}$) for different planetary ionospheres and ion species under varying ion-to-electron temperature ratios. Units are in V/m.}
\renewcommand{\arraystretch}{1.2}
\begin{tabular}{lcccr}
\\
\hline
Planet & Ion &  $\mathrm{T_i/T_e}=0.5$ & $\mathrm{T_i/T_e}=1$ & $\mathrm{T_i/T_e}=2$ \\
\hline
Venus & O$_2^+$  & 1.8246e-06 & 1.2553e-06 & 6.8599e-07 \\
Venus & O$^+$    & 8.3686e-07 & 5.1449e-07 & 1.9213e-07 \\
Earth & O$_2^+$  & 2.0339e-06 & 1.4123e-06 & 7.9067e-07 \\
Earth & O$^+$    & 9.4154e-07 & 5.9300e-07 & 2.4446e-07 \\
\hline
\label{tab:S5}
\end{tabular}
\end{table}

\begin{table}
\centering
\caption{Values of $E_{hf}$ at a fixed ionization degree ($\beta=10^{-4}$), electron temperature ($T_e=2000$~K) and electron temperature gradient ($\nabla T_e = 0.005~\text{K/m}$) for different planetary ionospheres and ion species under varying ion-to-electron temperature ratios. Units are in V/m.}
\renewcommand{\arraystretch}{1.2}
\begin{tabular}{lcccr}
\\
\hline
Planet & Ion &  $\mathrm{T_i/T_e}=0.5$ & $\mathrm{T_i/T_e}=1$ & $\mathrm{T_i/T_e}=2$ \\
\hline
Venus & O$_2^+$  & 2.2552e-06 & 1.9023e-06 & 1.5501e-06 \\
Venus & O$^+$    & 1.2695e-06 & 1.1637e-06 & 1.0582e-06 \\
Earth & O$_2^+$  & 2.4641e-06 & 2.0588e-06 & 1.6543e-06 \\
Earth & O$^+$    & 1.3739e-06 & 1.2420e-06 & 1.1103e-06 \\
\hline
\label{tab:S6}
\end{tabular}
\end{table}

\begin{table}
\centering
\caption{$E_{\text{hf}}$-to-$E_0$ ratio at a fixed ionization degree ($\beta=10^{-4}$), electron temperature ($T_e=2000$~K) and electron temperature gradient ($\nabla T_e = 0.01~\text{K/m}$) for different planetary ionospheres and ion species under varying ion-to-electron temperature ratios.}
\renewcommand{\arraystretch}{1.2}
\begin{tabular}{lcccr}
\\
\hline
Planet & Ion &  $\mathrm{T_i/T_e}=0.5$ & $\mathrm{T_i/T_e}=1$ & $\mathrm{T_i/T_e}=2$ \\
\hline
Venus & O$_2^+$  & 1.4935 & 2.1930 & 5.0871 \\
Venus & O$^+$    & 2.1952 & 5.0922 & -18.2969 \\
Earth & O$_2^+$  & 1.4388 & 2.0369 & 4.2509 \\
Earth & O$^+$    & 2.0390 & 4.2552 & -47.9387 \\
\hline
\label{tab:S7}
\end{tabular}
\end{table}

\begin{table}
\centering
\caption{Values of $E_0$ at a fixed ionization degree ($\beta=10^{-4}$), electron temperature ($T_e=2000$~K) and electron temperature gradient ($\nabla T_e = 0.01~\text{K/m}$) for different planetary ionospheres and ion species under varying ion-to-electron temperature ratios. Units are in V/m.}
\renewcommand{\arraystretch}{1.2}
\begin{tabular}{lcccr}
\\
\hline
Planet & Ion &  $\mathrm{T_i/T_e}=0.5$ & $\mathrm{T_i/T_e}=1$ & $\mathrm{T_i/T_e}=2$ \\
\hline
Venus & O$_2^+$  & 1.6853e-06 & 1.0464e-06 & 4.0746e-07 \\
Venus & O$^+$    & 6.9760e-07 & 3.0560e-07 & -8.6403e-08 \\
Earth & O$_2^+$  & 1.8947e-06 & 1.2034e-06 & 5.1214e-07 \\
Earth & O$^+$    & 8.0227e-07 & 3.8410e-07 & -3.4065e-08 \\
\hline
\label{tab:S8}
\end{tabular}
\end{table}

\begin{table}
\centering
\caption{Values of $E_{hf}$ at a fixed ionization degree ($\beta=10^{-4}$), electron temperature ($T_e=2000$~K) and electron temperature gradient ($\nabla T_e = 0.01~\text{K/m}$) for different planetary ionospheres and ion species under varying ion-to-electron temperature ratios. Units are in V/m.}
\renewcommand{\arraystretch}{1.2}
\begin{tabular}{lcccr}
\\
\hline
Planet & Ion &  $\mathrm{T_i/T_e}=0.5$ & $\mathrm{T_i/T_e}=1$ & $\mathrm{T_i/T_e}=2$ \\
\hline
Venus & O$_2^+$  & 2.5171e-06 & 2.2947e-06 & 2.0728e-06 \\
Venus & O$^+$    & 1.5314e-06 & 1.5562e-06 & 1.5809e-06 \\
Earth & O$_2^+$  & 2.7260e-06 & 2.4513e-06 & 2.1771e-06 \\
Earth & O$^+$    & 1.6358e-06 & 1.6344e-06 & 1.6330e-06 \\
\hline
\label{tab:S9}
\end{tabular}
\end{table}







\end{document}